\documentclass[twocolumn,showpacs,preprintnumbers,amsmath,amssymb]{revtex4}
\usepackage{graphicx}
\usepackage{float}
\usepackage[usenames,dvipsnames]{color}
\begin{document}
   \title{Comments on ``Remarks on the spherical scalar field halo in galaxies''}
   \author{Kung-Yi Su$^{1,3}$}
     \email{b95202049@ntu.edu.tw}
   \author{Pisin Chen$^{1,2,3,4}$}
     \email{pisinchen@phys.ntu.edu.tw}
     \affiliation{%
1. Department of Physics, National Taiwan University, Taipei, Taiwan 10617\\
2. Graduate Institute of Astrophysics, National Taiwan University, Taipei, Taiwan 10617\\
3. Leung Center for Cosmology and Particle Astrophysics, National Taiwan University, Taipei, Taiwan 10617\\
4. Kavli Institute for Particle Astrophysics and Cosmology, SLAC National Accelerator Laboratory, Stanford University, Stanford, CA 94305, U.S.A.
}%

   \date{\today}

\begin{abstract}
We comment on the general solution of the scalar field dark matter provided in the paper ``Remarks on the spherical scalar field halo in galaxies'' by Kamal K. Nandi, Ildar Valitov and Nail G. Migranov. The authors made a mistake in the general form of the tangential pressure profile $p_t(r)$, which deviates from the correct profile, especially when $r$ is small. Although this mistake does not alter significantly the value of $w(r)$ when the integration constant $D$ is small, we found that it does result in an overestimate of $w(r)$ when $D$ is large.
\end{abstract}

\maketitle

In the paper ``Remarks on the spherical scalar field halo in galaxies''\cite{Nandi}, the authors Kamal K. Nandi, Ildar Valitov and Nail G. Migranov generalized the solution of a scalar field dark matter model provided in the paper ``Spherical scalar field halo in galaxies'' published by Matos et al.\cite{Matos}. Under the assumption of constant rotation curve and spherical symmetry, Matos et al. proposed a consistent solution of the Einstein equation for static galactic halos as follows:
\begin{align}
ds^2=&-B(r)dt^2+A(r)dr^2+r^2d\theta^2+r^2sin^2\theta d\phi^2\notag\\
B(r)&=B_0r^{\ell}\notag\\
A(r)&=\frac{4-\ell^2}{4+D(4-\ell^2)r^{-(\ell+2)}}\notag\\
\Phi(r)&=\sqrt{\frac{\ell}{8\pi G}}\ln(r)+\Phi_0\notag\\
V(r)&=-\frac{1}{8\pi G(2-\ell)r^2},
\end{align}
where $\ell=(v_c/c)^2\ll 1$, $v_c$ is the constant rotation velocity, $\Phi$ is the scalar field and $V$ is its potential.
Under such solution, the resulting energy and density profiles are:
\begin{align}
 \rho(r)=-p_t(r)=\frac{-\ell^2}{(4-\ell^2)8\pi Gr^2}\notag\\
 p_r(r)=\frac{\ell(\ell+4)}{(4-\ell^2)8\pi Gr^2}.
\end{align}

 The solution of Matos et al. is a special solution with an integration constant fixed to 0, which Nandi et al. found to be unsuitable. Instead, Nandi et al. suggested that the general form of the energy and density profiles should be as follows:
\begin{align}
\rho(r)&=\frac{1}{8\pi G}\frac{r^{-(4+\ell)}[D(\ell^3+\ell^2-4\ell-4)+\ell^2r^{2+\ell}]}{\ell^2-4}\notag\\
p_r(r)&=\frac{1}{8\pi G}\frac{r^{-(4+\ell)}[D(\ell^3+\ell^2-4\ell-4)-\ell(4+\ell)r^{2+\ell}]}{\ell^2-4}\notag\\
p_t(r)&=\frac{1}{8\pi G}\frac{r^{-(6+\ell)}[D(\ell^3+\ell^2-4\ell-4)+\ell^2r^{2+\ell}]}{4(\ell^2-4)}\notag\\
&\times[(r^2-1)\ell-2(r^2+1)]
\end{align}
where $D$ is an integration constant. They further argued that $D=0$ would induce a result that violates the weak energy condition, which is solvable if $D$ is set to be larger than $10^{-7}$. According to them, once $D>10^{-7}$, $w\equiv (p_r+2p_t)/(3\rho)$ will be larger than $-1$ and $\rho$ will be larger that $0$ for the entire range of galatocentric radius $r$ of interest.

However, we found the general solution proposed by Nandi et al. curious. The first thing to be mentioned is that the energy momentum of a scalar field reads:
\begin{align}
T_{\mu\nu}=\Phi_{,\mu}\Phi_{,\nu}-\frac{1}{2}g_{\mu\nu}\Phi^{,\sigma}_{,\sigma}-g_{\mu\nu}V.
\end{align}
Therefore, if all relevant physical quantities are only dependent on $r$, which is the case considered by both Matos et al. and Nandi et al., then one should have the following relations,
\begin{align}
\rho(r)=-T^0_0=\frac{1}{2}\Phi^{,r}\Phi_{,r}+V\notag\\
p_r(r)=T^r_r=\frac{1}{2}\Phi^{,r}\Phi_{,r}-V\notag\\
p_t(r)=T^\theta_\theta=T^\phi_\phi=-\frac{1}{2}\Phi^{,r}\Phi_{,r}-V,
\end{align}
which means that $\rho=-p_t$ is generically true. However, the general solution provided by Nandi et al. does not honor this relationship. Furthermore, their $p_t(r)$ does not recover the solution of Matos et al. when $D=0$. We accordingly believe that the profile of tangential pressure of Nandi et al. is incorrect.

To address this issue, we resort to numerical solution of the Einstein equation. We found that both $p_r(r)$ and $\rho(r)$ resulted from the numerical calculation agree with the solution of Nandi et al. without any visible difference, while our $p_t(r)$ differs from that of Nandi et al. in that theirs saturates to half of our value. We have tried a wide range of both $D$ and $\ell$, which enabled us to claim that such a difference is general. In fact, such outcome is expected since their density and tangential pressure profiles follow the relationship $\rho(r)\times[(r^2-1)\ell-2(r^2+1)]/(4r^2)=p_t(r)$. In the reality, $\ell$ is of the order $10^{-6}$, which is so small that $[(r^2-1)\ell-2(r^2+1)]/(4r^2)\approx -2(r^2+1)/(4r^2)$. Therefore when $r$ is large, their $p_t(r)/\rho(r)=-(1/2)$, which is exactly half of what it should be. The comparison between $p_t(r)$ and $2p_t(r)$ provided by Nandi et al. as well as the $p_t(r)$ resulting form the correct numerical calculation are shown in Fig. 1, from which we see that when $r$ is small their $p_t(r)$ is significantly deviated from the correct one, while their $2p_t(r)$ saturates at large $r$ to the correct value.
\begin{figure}[h]
  \includegraphics[width=7.3cm]{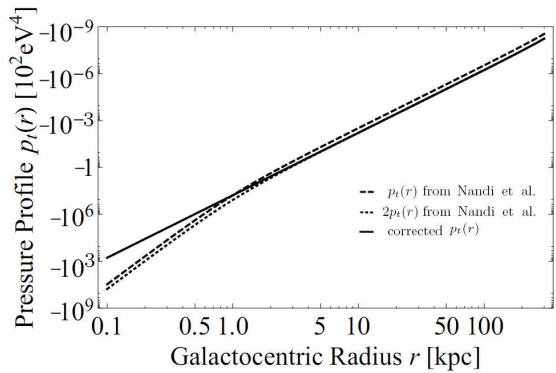}\\
  \caption{The comparison of the $p_t(r)$ and $2p_t(r)$ from Nandi et al. as well as the correct $p_t(r)$ from numerical calculation. The dashed curve and the dotted curve is their $p_t(r)$ and $2p_t(r)$ respectively. The solid curve is the corrected $p_t(r)$}
\end{figure}

With regard to the equation of state $w(r)$ parameter, we note that $\ell$ is in general an extremely small value. Accordingly, if $D$ is comparable to $\ell$, we can see from Eq. 3 that $p_r(r)$ is dominated by the term without $D$, whereas both $p_t(r)$ and $\rho(r)$ are dominated by the $D$ term. However, since the dominating term in $p_r(r)$ is larger than that of either $p_t(r)$ or $\rho(r)$ when $r$ is not too small, $p_r(r)$ is several orders of magnitude larger that $\rho(r)$. Therefore, while fixing the lower bound of $D$, the realm considered is in general less than $10^{-6}$, which renders the mistake in $p_t(r)$ too small to be relevant.

On the contrary, the situation is different at large $D$, since in this case all three profiles are dominated by the $D$ term, which is identical for all. The comparison between the $w(r)$ based on Nandi et al. and the correct $w(r)$ under two different $D$ values is shown in Fig. 2. We found that when $D$ becomes larger, their $w(r)$ grows from 0 slower while our $w(r)$ grows from $-1/3$ slower, instead.
\begin{figure}[h]
  \includegraphics[width=7.3cm]{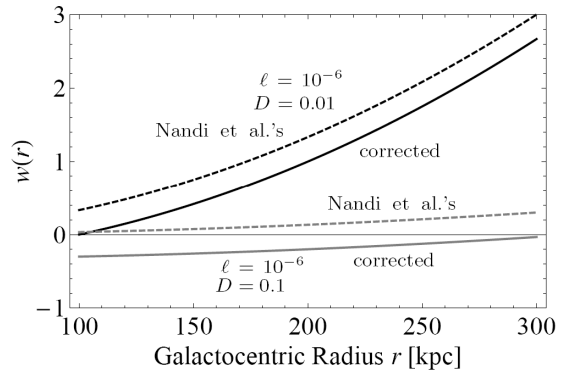}\\
  \caption{The comparison of Nandi et al.'s $w(r)$ and the corrected $w(r)$ when $D$ is larger. The solid curves are the corrected result while the dashed ones are theirs. The black curves correspond to the situation with $\ell=10^{-6}$ and $D=0.01$. The gray curves correspond to the situation with $\ell=10^{-6}$ and $D=0.1$.}
 \end{figure}

In conclusion, the general expression of $p_t(r)$ for scalar field dark matter provided by Nandi et al. should be corrected. Since both $\rho(r)$ and $p_r(r)$ are correct, we suggest that the correct form of $p_t(r)$ should be
\begin{align}
 p_t(r)&=-\frac{1}{8\pi G}\frac{r^{-(4+\ell)}[D(\ell^3+\ell^2-4\ell-4)+\ell^2r^{2+\ell}]}{\ell^2-4}\notag\\
 &\equiv-\rho(r)
\end{align}
  Although such correction will not affect the value of $w(r)$ and, accordingly, the lower bound of $D$, when $D$ is small, it still leads to a sizable change of $w(r)$ when $D$ is large. 

\section*{Acknowledgement}
We thank S.H. Shao, Y. D. Huang and C. I. Chiang for interesting and inspiring discussions. This research is supported by Taiwan National Science Council under
Project No. NSC 97-2112-M-002-026-MY3 and by US Department of Energy under Contract No. DE-AC03- 76SF00515. We also thank the support of the National Center for Theoretical Sciences of Taiwan.

\end{document}